# Role of ambient air on photoluminescence and electrical conductivity of assembly of ZnO Nanoparticles

Manoranjan Ghosh,[a),#] R. S. Ningthoujam,[b)] R.K. Vatsa,[b)] D. Das,[b)] V. Nataraju,[a)] S.C. Gadkari,[a)] S. K. Gupta,[a)] and D. Bahadur[c)]

*Bhabha Atomic Research Centre, Mumbai-400085, India*

**Abstract:** Effect of ambient gases on photoluminescence (PL) and electrical conductivity of films prepared using ZnO nanoparticles (NPs) have been investigated. It is observed that NPs of size below 20 nm kept inside a chamber exhibit complete reduction in their visible PL when oxygen partial pressure of the surrounding gases is decreased by evacuation. However the visible PL from ZnO NPs is insensitive to other major gases present in the ambient air. The rate of change of PL intensity with pressure is inversely proportional to the ambient air pressure and increases when particle size decreases due to the enhanced surface to volume ratio. On the other hand an assembly of ZnO NPs behaves as a complete insulator in the presence of dry air and its major components like $N_2$, $O_2$ and $CO_2$. Electrical conduction having resistivity $\sim 10^2 - 10^3$ $\Omega$m is observed in the presence of humid air. The depletion layer formed at the NP surface after acquiring donor electrons of ZnO by the adsorbed oxygen, has been found to control the visible PL and increases the contact potential barrier between the NPs which in turn enhances the resistance of the film.

---

[a)] Technical Physics Division
[b)] Chemistry Division
[c)] Indian Institute of Technology, Bombay, Mumbai – 400070, India
[#] Email: mghosh@barc.gov.in

## I. INTRODUCTION

ZnO has emerged as one of the most researched wide band gap (3.3 eV) optoelectronic materials due to its attractive optical properties.[1,2] It is cheap, nontoxic and widely used in industries related to rubber, paint and healthcare. Renewed interest on ZnO has been observed due to its potential applications in the area of light emitting devices,[3] spintronics,[4] transparent conducting film[5] and sensors. ZnO based sensors are most promising and widely investigated due to high sensitivity to different chemical environments.[6-17] The non-toxic character of ZnO makes it a suitable candidate as a biosensor as well.[6] In particular, porous film of chemically synthesized ZnO nanostructures having a high surface to volume ratio provides large interaction area for the adsorbent on the surface.

ZnO nanoparticles (NPs) are very sensitive to the surrounding liquid as well as gaseous environment. The optical properties of suspended ZnO NPs can be controlled by using solvents having different electrochemical properties.[7] ZnO based single crystal, ceramic as well as assembly of NPs have been investigated for sensing various harmful inorganic gases such as $H_2S$, $NO_2$, $Cl_2$, CO, and $H_2$.[8-11] ZnO has emerged as a promising ethanol sensor,[12,13] and possibility of sensing other organic gases/vapors such as acetone, toluene, methanol, $CH_4$, acetyline is now an active field of research.[14,15] Due its fast response and recovery times, ZnO is also considered as an excellent sensor of humidity which is omnipresent in the atmosphere.[16,17] Thus ZnO can sense multiple gases and cautious steps are necessary for designing ZnO based sensor for selective detection of gases. The films made from ZnO NPs show high resistance (~ 1 MΩ) in the presence of wet air and become non-conducting under major components of air. Therefore, the effect of air needs to be excluded while studying sensing properties of other toxic gases. On similar ground, it is important to understand the electrical conduction and charge transport mechanism in films made by assembly of ZnO NPs.

Most of the studies on gas sensing properties of ZnO are conductometry based, i.e. the effect of adsorbed gases on electrical conductivity of the sensor element is measured. In search for better selectivity, response and sensitivity, gas sensing based on alternative properties of ZnO needs to be explored. NPs of ZnO show characteristic near-band-edge (NBE) emission in the ultra violet (UV) region (~380 nm) due to excitonic recombination and surface related visible photoluminescence (PL) in the wavelength range of 450–650 nm.[18] The visible PL originates from defects such as charged oxygen vacancies that are believed to be located near the surface.[19] The intensity of this broad emission is highly sensitive to the environment and mainly depends on the surface to volume ratio of the NPs.[7,19] Optical technique based on this visible emission of ZnO can be used for the purpose of sensing gases. Adsorption of ppm level of $NO_2$ has been detected by change in the visible PL from nanobelts of $SnO_2$, a wide band gap semiconductor like ZnO.[20] Effect of ambient gases on photoluminescence of ZnSe nanowires has been investigated.[21] In the present investigation, we explore the possibilities of sensing gases by ZnO NPs through monitoring its optical emission as a function of ambient gas pressure. We demonstrate that ZnO NPs of size less than around 20 nm exhibit drastic fall in their visible PL intensity when oxygen pressure inside a chamber is decreased. However, other gases in the atmosphere do not affect the optical emission from ZnO NPs. The oxygen partial pressure dependent surface luminescence from ZnO NPs demonstrated here has not been investigated earlier. Due to the surface related origin of this broad visible PL, the sensitivity can be easily tuned by changing the particle size. The observation is important for determination of the oxygen partial pressure in the air which is a biologically relevant quantity. Finally, the oxygen pressure dependent visible PL demonstrated in this work establishes a direct link between charged oxygen vacancies and visible PL of ZnO NPs.

## II. SAMPLE PREPARATION AND EXPERIMENTAL TECHNIQUES

ZnO NPs have been synthesized by dissociation of $Zn(CH_3COO)_2 \cdot 2H_2O$ in a basic medium at temperatures 60-75°C. For synthesizing nanoparticles of size 10 nm, $0.03M$ NaOH solution in ethanol is slowly added in $0.01M$ solution of $Zn(CH_3COO)_2 \cdot 2H_2O$ in ethanol kept at 65 °C. The final solution was stirred and heated at 65 °C for 1 h. This method allows precipitation of ZnO nanoparticles and avoids precipitation of hydroxides if the temperature is more than 60°C. Precursor concentrations, reaction time and temperature have been varied to synthesize particles of different sizes. For obtaining 5 nm particles, 0.005M of $Zn(CH3COO)_2 \cdot 2H_2O$ was reacted at temperature 60°C for 30 min. 0.015 M of $Zn(CH_3COO)_2 \cdot 2H_2O$ has been reacted at temperature 75°C for 6 h to synthesize particles of size 15 nm.

The nanostructured ZnO film (thickness varying from 1-1.5 μm) is obtained by adding nanoparticle dispersion in ethanol drop wise on a pre-decided area of a commercially available indium-tin oxide (ITO) coated glass or quartz substrate. The ZnO dispersion has been spread over the surface uniformly and dried at room temperature. Equal number of layers was added to obtain films of similar thickness repeatedly. To assure the presence of a ZnO layer all over the substrate, a thick film is deposited.

The transmission electron microscopy (TEM) images of nanoparticles were collected by a JEOL-HRTEM, operated at 200 keV. The crystal structures and size of all the samples have been analyzed by an X-ray diffraction technique using a Philips X'pert-Pro Diffractometer. Thickness of the film (1-1.5 μm) is obtained from the VECCO Atomic Force Microscopy image of a scratched ZnO film. An optical cryostat equipped with the provision of evacuation and passing select gases is used for PL measurements by a Spectrofluorimeter from Edinburg Instruments. For electrical measurements, gold contact pads of 2 mm x 2 mm area having 100 μm separations have been deposited. Current vs. time measurement at fixed voltages has been performed by a Keithley 6487 picoammeter/voltage source connected with the sample by copper wires using silver paint.

## III. RESULTS

### A. Characterization by TEM, XRD, SEM and AFM

The representative TEM images of collection of particles of average sizes 5, 10 and 15 nm are shown in Fig. 1(a), (b) and (c) respectively. HRTEM images (data not shown) confirm the crystalline nature of the individual NP.[22] The X-ray diffraction results are plotted in Fig. 1(d) for samples having sizes 5, 10 and 15 nm. The indexed X-ray diffraction pattern confirms Wurtzite symmetry of the synthesized ZnO nanostructures. Average sizes of the nanostructures are estimated by employing Williamson-Hall analysis method by which peak broadening due to strain can be excluded.[22,25] The size determined by TEM and XRD analysis agrees well within a range of ±2 nm. The root mean square (rms) distribution in size ($d$) can be quantified as $(d_{rms}/d)$

< 10% for all the samples investigated. Interestingly ZnO cluster size in the film prepared by drop casting of suspended ZnO NPs becomes 40-50 nm due to agglomeration [Fig. 2(a)]. Thickness of the film (1-1.5 μm) is obtained from the Atomic Force Microscopy (AFM) image of a scratched ZnO film [Fig. 2(b)].

**B. Photoluminescence as a function of ambient air and oxygen pressure**

The phenomenon of air and oxygen pressure dependent visible PL is investigated on ZnO NPs spread out as a film. Spherical NPs of sizes 5, 10 and 15 nm have been chosen for demonstration purpose. First we recorded the PL spectra of ZnO NPs (dia~15nm) deposited on quartz substrates kept inside the optical cryostat shown in a diagram (Fig. 3). For better visibility of NBE and visible bands, the film of ZnO NPs in this study is excited at 325 nm. The sample chamber was initially at ambient air atmosphere. When evacuation starts the chamber pressure decreases, thereby reducing the visible PL intensity in the range of 450-650 nm which shows a sharp dependence within the tested range of chamber pressure (Fig. 4). Almost complete reduction (95%) of visible PL was observed at the highest vacuum level ($4.6\times10^{-5}$ mbar) that could be achieved with this chamber. Therefore, visible emission from ZnO NPs can be very effectively controlled by changing the ambient air pressure. Hence, the pressure inside the chamber can be estimated by measuring PL intensity of ZnO NPs from a calibrated pressure dependent intensity curve.

Since air is a mixture of several gases (primarily containing $N_2$ and $O_2$) individual gases have been tested for their role in the phenomena described above. In Fig. 5(a), we show the PL spectra of the film of ZnO NPs (dia~10 nm) kept in different gas environments. If the evacuated chamber is filled with oxygen, the PL intensity shoots up immediately (within 30 seconds) to 1.5 times of the value obtained in air atmosphere. This is because the chamber filled with oxygen has higher oxygen partial presure than that of air. Therefore oxygen partial pressure can be effectively sensed by the film made of ZnO NPs. However PL intensity of the evacuated chamber does not recover if it is filled by $N_2$ and He seperately. Therefore the visible PL intensity from ZnO is insensitive to the gases like $N_2$, $H_2$, He, and Ar.

Once the decrease in the partial pressure of oxygen is identified as the responsible candidate for reduction in the emission intensity, we systematically collected PL spectra at various oxygen pressures. As shown in Fig. 5(b), the intensity decreases as the oxygen pressure decreases as in the case of air. In Fig. 6, the integrated intensities corresponding to logarithm of various ambient pressures have been plotted for different particle sizes. From the graph it can be concluded that the rate of change of PL intensity with pressure (or the sensitivity) is inversely proportional to the ambient air pressure. i.e.,

$$\frac{dI}{dP} \propto \frac{1}{P} \qquad (1)$$

Or in other words, the intensity of the visible PL is proportional to the logarithm of the ambient chamber pressure (depicted by linear fitting of the data points in Fig. 6). This establishes the fact that the films of ZnO NPs investigated in this work are most suitable as a low pressure sensor.

It would be interesting to compare the sensitivity or rate of change of intensity due to pressure for various situation discussed above. From Fig. 6, it can be clearly seen that the sensitivity increases with the reduction in particle size. Therefore monitoring of the oxygen partial pressure by change of visible PL is a surface related phenomenon and depends on the surface to volume ratio of the NPs. The sensitivity (slope of linear fit) shows a size (1/diameter) dependence as expected. Further the sensitivity in pure oxygen environment is slightly higher than that in the air. This is because the rate of decrease in partial pressure of oxygen in air during evacuation is slower than the situation when pure oxygen is considered. The range of oxygen partial pressure which can be sensed by this optical method depends clearly on the particle size. The range increases with the decrease in particle size at the cost of their sensitivity. The visible PL vanishes quickly for lower particle sizes. Therefore the sensitivity as well as the sensing range of oxygen partial pressure can be optimized as per the requirement by simply varying the size of the NPs.

The sensing range of oxygen partial pressure also depends on the thickness of the ZnO films up to a certain thickness. The visible PL vanishes within 8 mbar, 0.45 mbar and 0.31 mbar for NP film of thickness ~20, ~40 and ~80 nm, respectively. There is no variation in the sensing range beyond the thickness of ~100 nm. The film thicknesses were kept within the range of 1-1.5 μm for all other samples investigated in this work. The sensitivity however does not show much variation with the thickness of the NP film.

**C. Influence of different gases on electrical conductivity**

We have seen that the visible emission from ZnO NPs showed a systematic dependence on the partial pressure of oxygen in the air environment. However the electrical resistance of the film remains unaffected due to the change in the ambient oxygen pressure. In case of NP assembly, transport of electrons occurs through the individual NPs at the surface connected in a network. Therefore the contact barrier between the NPs decides the resistance of the sensor element and controls the charge transport. The change in resistance is caused by adsorption or desorption of gases at the surface. The film made of ZnO NPs show very high resistivity ($10^2 – 10^3$ Ωm) in air. To check the effect of individual gases present in air, current through the sensor element (film of ZnO NPs of size ~ 10 nm) was measured during the evcuation of the chamber. The current decreases rapidly when evacuation takes place as shown in Fig. 7. Under vacuum level of mere 1 mbar, the film behaves like a complete insulator due to the high contact barriers between the nanoparticles (discussed later). For the purpose of introducing other gasses, the chamber was evacuated up to the level of ~$10^{-5}$ mbar. Then individual gases like $O_2$, $N_2$, $H_2$ and $CO_2$, were introduced. To avoid mixing of gases, the chamber was evacuated down to ~$10^{-5}$ mbar between the applications of two consecutive gases. Interestingly, the resistance of the assembly of ZnO NPs is insensitive to the major constituents of air such as $N_2$, $O_2$ and $CO_2$ (Fig. 7). Even, the application of dry air was not found enough to regain the conductivity recorded initially when the sensor element was kept in the atmospheric air. Certain percentage of conductivity has been recovered only after the application of humid air (RH 90%). Therefore, the film of ZnO nanoparticles kept in air show electrical conduction only due to the presence of moisture in the air (discussed later). It should be noted that similar kind of film made by assembly of ZnO nanoparticles exhibits p-type conductivity and respond accordingly on application of strong oxidizing ($Cl_2$) and reducing ($H_2S$) gases in the air environment.[9] Conductometric response of ZnO NPs to the toxic inorganic gases and organic vapours are beyond the scope of the paper and hence not included in this section.

## IV. DISCUSSIONS

The effect of air environment on PL and electrical conductivity as described above is decided by two factors. NPs in an assembly are connected to each other through grain boundaries. So, there exist high contact potential barrier between the NPs and hence behaves as an insulator. Also the NPs are in contact with the air environment. Therefore exchange of electron is possible and after equilibration a depletion layer will form at the surface of NPs. Width of the depletion layer is high for NPs of size ~ 10 nm which severely affect the optical emission from ZnO (discussed in the next subsection). The surface depletion layer can also enhance or reduce the contact potential barrier. But for nanoparticles assembly having large number of junctions is mainly contact controlled.[12] The modification of the surface depletion and the contact potential due to the ambient air medium is discussed in the subsequent sections.

### A. Mechanism of oxygen pressure dependent visible luminescence

In line of the earlier studies we propose that the visible PL from ZnO NPs originates from the charged oxygen vacancies that are predominantly located near the surface.[23] Free-carrier depletion at the particle surface and its effect on the ionization state of the oxygen vacancy has strong impact on the visible emission.[24] An oxygen vacancy created in a ZnO NP is neutral when not in contact with the surrounding fluid medium. The donor electrons in the conduction band of neutral oxygen deficient ZnO will occupy the acceptor like surface states created by excellent acceptor oxygen in atmosphere. By this way singly ($V_o^+$) and doubly ($V_o^{++}$) charged oxygen vacancies are created and the surface becomes positively charged. The existence of the positively charged depletion layer can be directly seen from the zeta potential (~18 mV) measured in these NP of small size dispersed in ethanol.[7,25] It can be shown that the broad visible emission in the blue-yellow region is a composite of two lines located approximately at 2.2 eV (550 nm) and at 2.5 eV (500 nm).[26] Emission band appearing around 550 nm has been suggested to originate from $V_o^{++}$ whereas $V_o^+$ is responsible for the emission band around 500 nm [Fig. 8(a)].[25,26] The relative intensities and positions of these two lines determine the overall nature of the visible emission. The transfer of donor electrons to the acceptor like surface states forms a depletion layer near the surface [Fig. 8(a)]. Such a depletion layer gives rise to band bending in nanospheres whose size is comparable to the depletion width (~5–10 nm).[27] The extent of band bending in the near surface depletion layer in the NP is linked to their surface charge which decides the predominant nature of the visible emission. For positive charge on the surface, the band bending changes the chemical potential, which populates the level $Vo^{++}$ preferentially compared to the other level, leading to an overall enhancement in the visible PL intensity. When oxygen is fully evacuated, no more electron transfer is possible. The ZnO NP having charged oxygen vacancies at the surface will be neutral ($Vo^X$).[26,28] This modification of the surface charge by oxygen evacuation will change the band bending in such a way that there will be a gradual reduction in the visible PL. Depending on the NP size, at sufficiently low oxygen pressures ZnO NP having neutral oxygen vacancy does not show the surface related visible emission.

**B. Electrical conduction in presence of air**

As mentioned earlier, there exist a large number of junctions between the electrodes in case of NPs connected in a network. The charge transport occurs through a grain boundary that offers a high potential barrier for electrons to flow. The contact barrier with varying thickness and height determines the transport of electrons through the NPs. The resistance of such a film is given by, [12,13]

$$R = R_0 \exp[-\frac{e(V_0 - V_g)}{k_B T}] \qquad (2)$$

Where $\Delta V = (V_0 - V_g)$ is the change of the potential barrier in presence of any gas ($V_g$) from the contact potential in vacuum ($V_0$). $R_o$ is a factor including the resistance in vacuum and other parameters; e is the electronic charge, $k_B$ Boltzmann's constant and T absolute temperature. $\Delta V$ can be modified by the exposure of different gases and become substantial for nanoparticles having a high surface to volume ratio. In absence of gas adsorbent (i.e., under vacuum) ZnO NPs considered here behave as an insulator. Therefore the height $V_0$ is such that it completely blocks the electron transport through the junctions between the NPs. In presence of ambient gases a depletion layer forms at the surface. The gas molecules adsorbed in the pores, induce band bending to the associated NPs such that the contact potential can be modified depending on the nature of the gas molecules. Extent of band bending as well as the modulation of the depletion width is high for NPs having high surface to volume ratios and surface depletion becomes a dominant factor to control the resistance of the sensor element.

After exposure, oxygen is adsorbed on the surface and pores of the film by sharing donor electrons from the ZnO NPs. Negatively charged surface oxygen introduces upward band bending and forms a depletion layer. It reduces the conducting width across the electrode and increases the potential barrier of the contacts between the particles and therefore causes a drop in the conductivity. In Fig. 8(b), a band diagram of two NPs, holding $O_2$ molecules within the pore has been depicted. The diagram is union of the situation presented in Fig. 8(a), that describes the junction between the nanoparticle and infinite air medium. If the depletion width is significantly smaller than the diameter of the NP, then surface depletion has little impact on the density and mobility of the electrons. But potential barrier is influenced by this surface depletion layer as shown in Fig. 8. As a result the contact potential in presence of adsorbed oxygen is much higher than that in vacuum. Due to which resistance of the film increases further and behaves like insulator as in the case of vacuum. Other major gases in the atmosphere such as $N_2$, $CO_2$, are not known for donating or accepting electrons from the NPs thus no further change in the conductivity is observed.

One of the principal motivations of this investigation is to find out the responsible candidate for electrical conduction of the nanoparticle film. We identified that moisture omnipresent in the air atmosphere significantly increases the conductivity of the film. ZnO is considered as an excellent humidity sensor because of its high sensitivity and fast response and recovery times.[16,17] When dry air is incorporated in the evacuated chamber no change in conductivity was observed (Fig. 7). The current rises to a measurable limit (0.1 – 1 µA) only after the introduction of air from the atmosphere. Therefore the fast (within 2-3 sec) response observed after introduction of wet air (RH 90%) is due to the water vapour present in the atmosphere. Water can be adsorbed in vapour phase as well as may condense on the surface. For low humidity, water molecules may be adsorbed on the surface in the vapour phase. In this case, possibilities of electrolytic conduction can be ruled out due to the existence of discontinuity in the water layer created at the surface.

Rather dissociation of water is more probable due to various factors as discussed here. Oxygen vacancies are known as active sites for water dissociation.[29] Also the oxygen ion adsorbed on the surface can dissociate water molecule through the transfer of one proton to a nearby oxygen atom.[30] Therefore two hydroxyl groups are formed for every oxygen atom. Further, auto-dissociation of the water molecules even at low temperatures is allowed because of the small dissociation barrier.[29-32] Moreover, the increased roughness present in the drop cast film enhances the surface reactivity as well as water dissociation by creating high local charge density and a strong electrostatic field at the top of the nanoparticles.[16,29-32] Dissociation of water provides protons as charge carriers for the hopping transport that reduces the resistance of the film. At high humidity however one can expect electrolytic conduction along with protonic transport due to the formation of multiple water layers.

## V. CONCLUSIONS

An in depth investigation on the role of ambient air on the photoluminescence and electrical conduction of the ZnO NPs has been documented. The phenomenon of reduction of the visible emission by lowering oxygen pressure opens up a way to exploit the potential of ZnO NPs in low oxygen pressure measurement through an optical technique. The sensitivity can be improved by simply using smaller particles having higher intensity of the surface related emission band. Electrical conduction in porous film made by assembly of ZnO NPs is possible due the presence of moisture in the air. A direct link between the charged oxygen vacancies and the visible PL demonstrated here may resolve the long standing debate on the origin of surface defect related emission from ZnO NPs.

## ACKNOWLEDGMENTS

One of the authors, MG is thankful to the Department of Science and Technology for financial support as DST NANO Mission.

**FIGURE CAPTIONS**

FIG. 1. TEM images of collection of nanoparticles of average sizes (a) 5 nm, (b) 10 nm and (c) 15 nm. Indexed XRD patterns of these nanoparticles as indicated on the graph show wurtzite structure [(d)].

FIG. 2. (Color online) (a) The film of ZnO nanoparticles (size ~15 nm) when coated on substrate shows agglomerated cluster of size 40 nm. (b) AFM image of scratched film gives film thickness ~1.2 μm.

FIG. 3. (Color online) Experimental arrangement adopted for gas pressure dependent photoluminescence measurement.

FIG. 4. (Color online) Photoluminescence spectra after excitation at 325 nm from a film of ZnO nanoparticles of size~15 nm for different air pressures.

FIG. 5. (Color online) Photoluminescence spectra from ZnO nanoparticles of size~10 nm (a) in different environments and (b) for different oxygen pressures as indicated on the graph.

FIG. 6. (Color online) Integrated luminescence intensity shows approximately linear dependence with ambient pressure. Sensitivity increases when particle size decreases at the cost of sensing range of oxygen pressure.

FIG. 7. The film of ZnO NPs of size 10 nm is used for sensing ambient air and its major component gases by conductometric measurement. Vacuum level of $10^{-5}$ mbar is created in between the application of two different gases.

FIG. 8. (a) Energy band diagram of optical emission based oxygen sensing. Visible emission (combination of lines 2.2 and 2.5 eV) from ZnO can be tuned by band bending in the depletion region formed due to adsorbed oxygen on the surface. (b) A model describing the junction between two nanoparticles having adsorbed oxygen in the pores. Upward band bending due to oxygen enhances the contact potential between the nanoparticles.

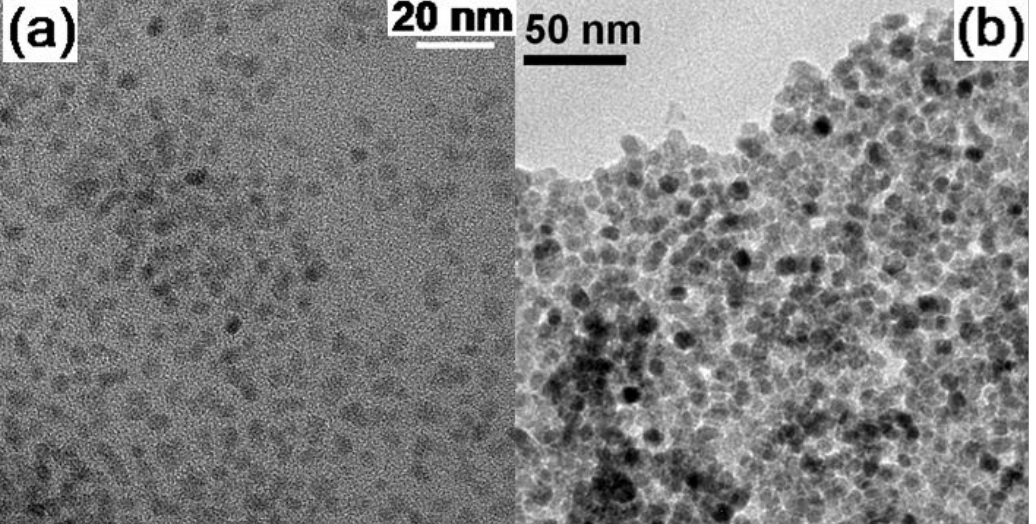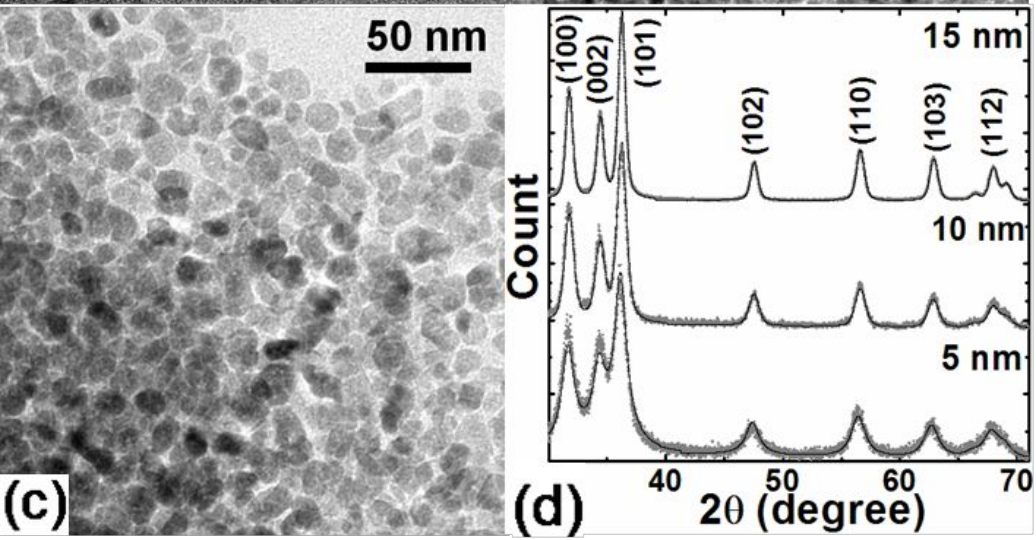

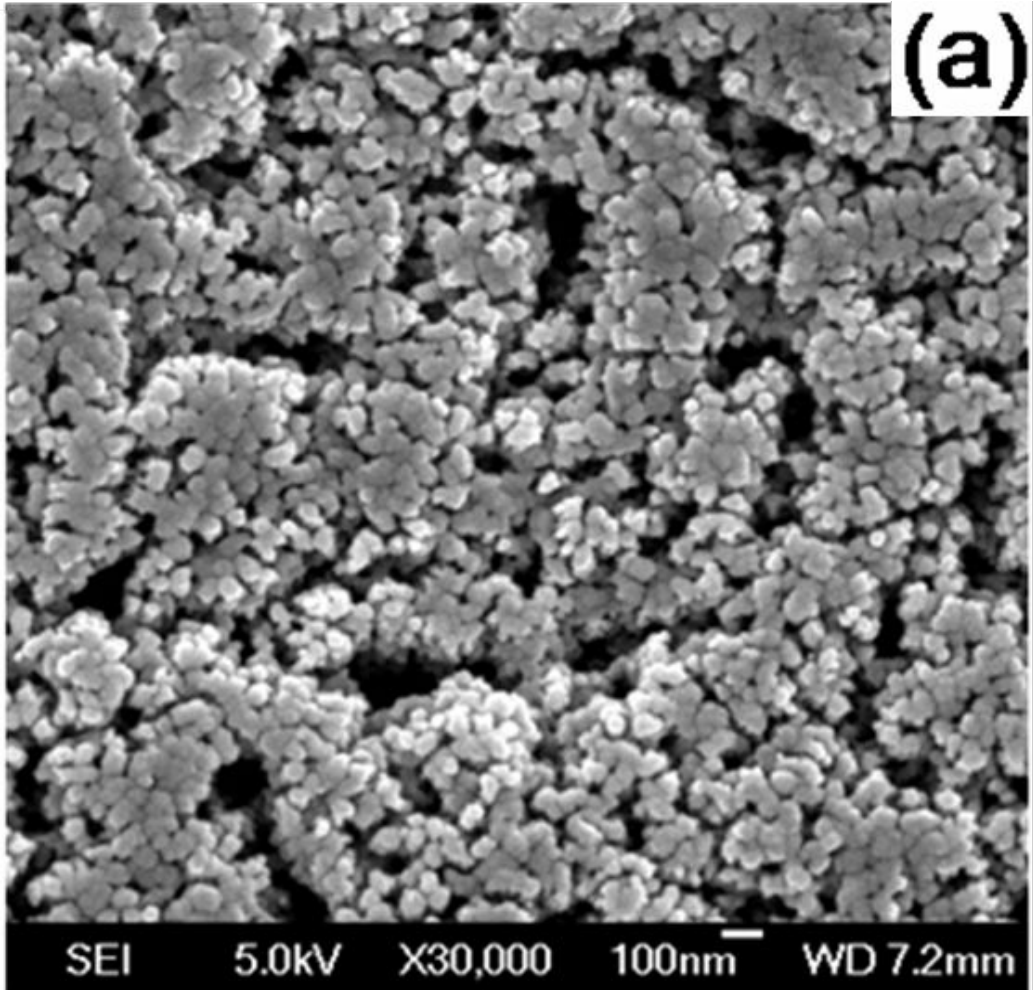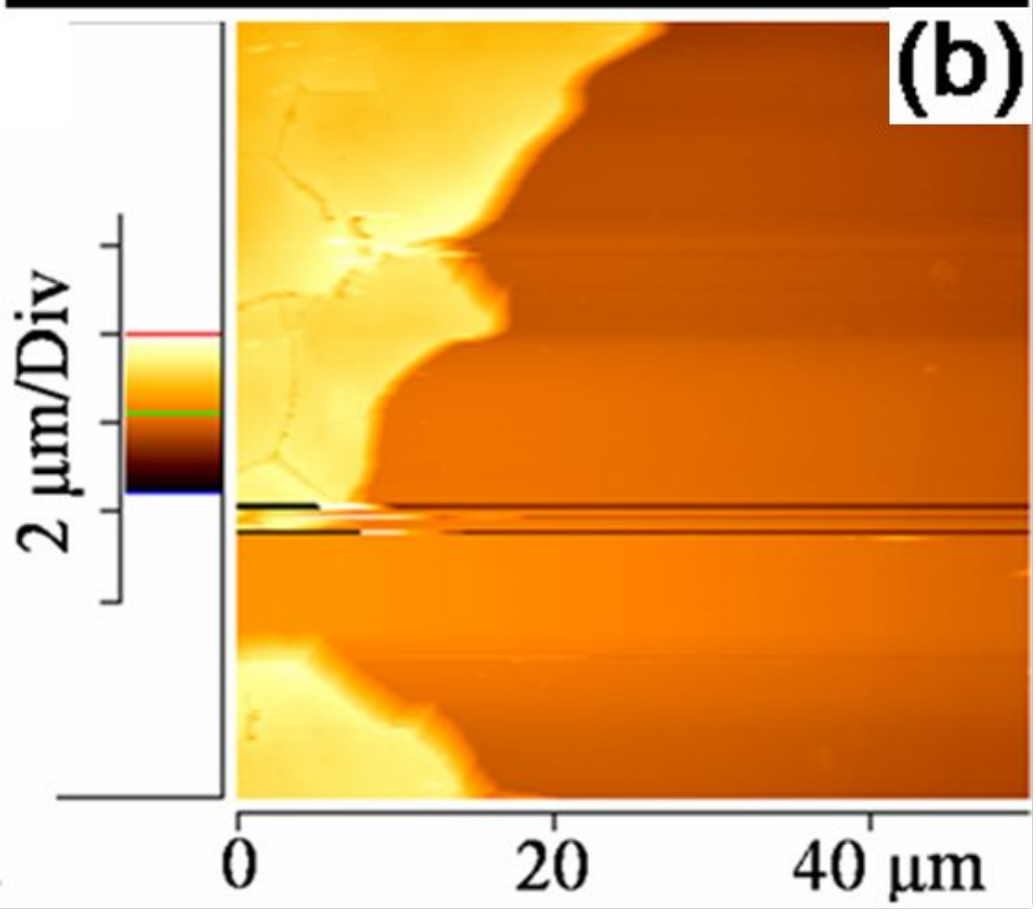

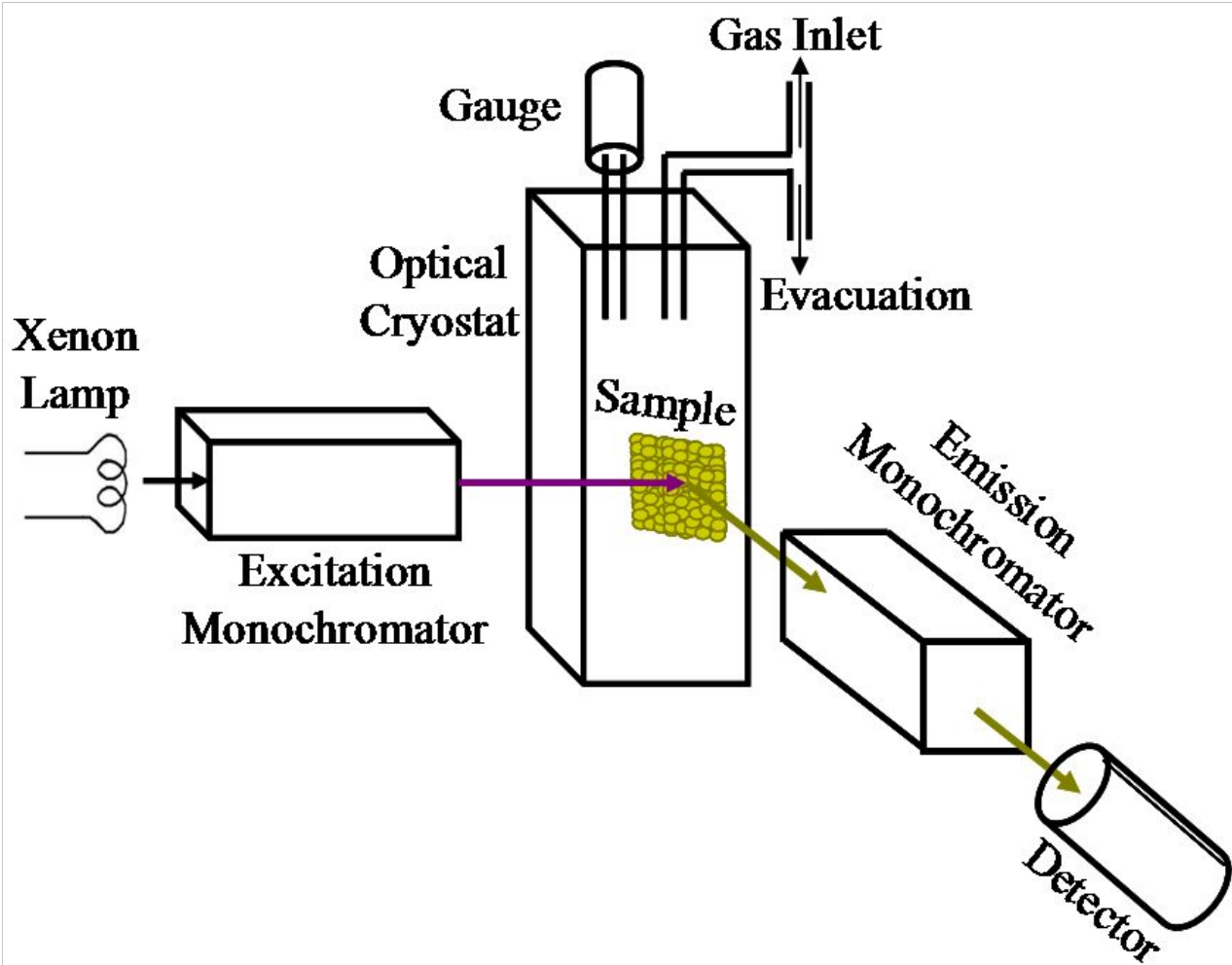

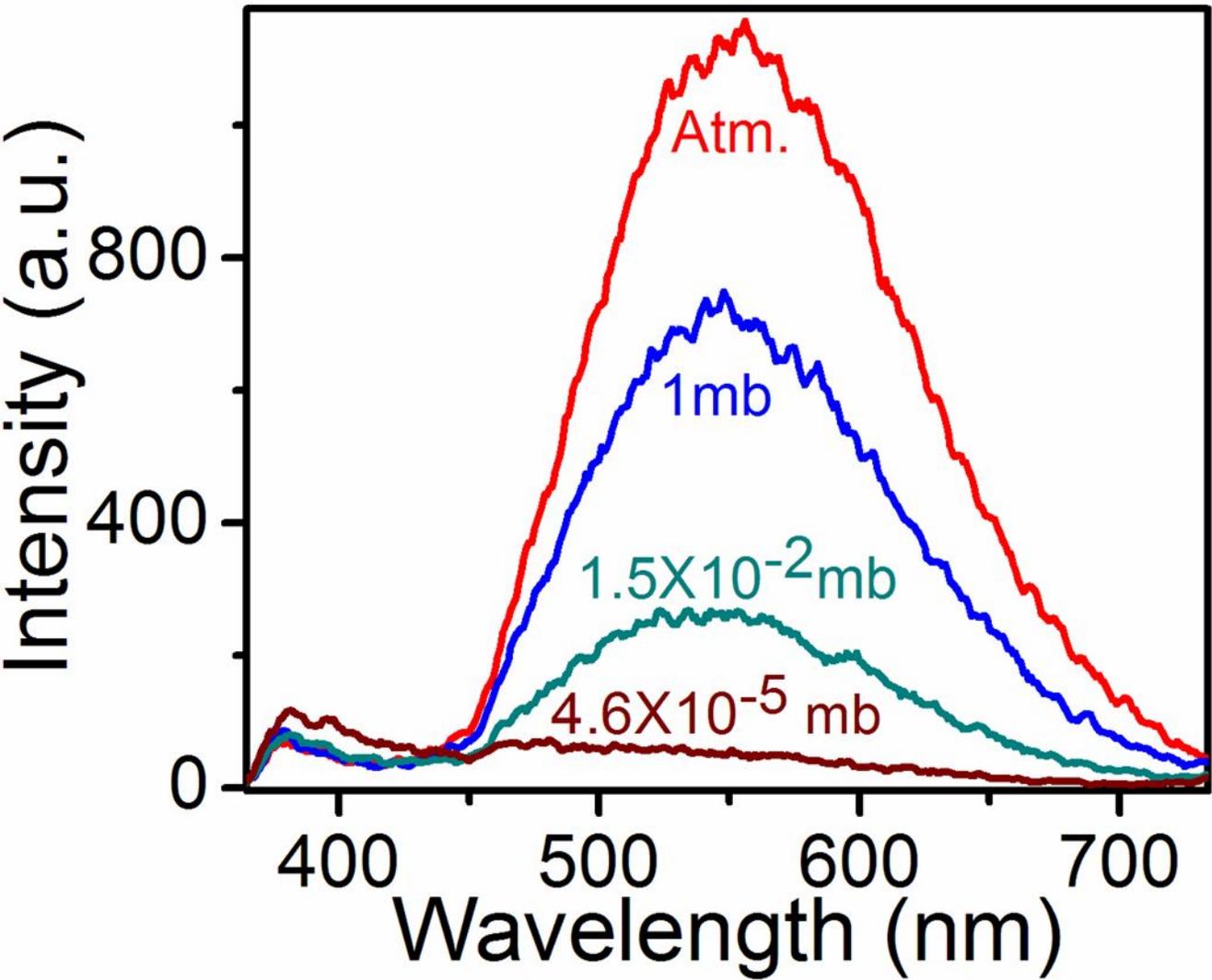

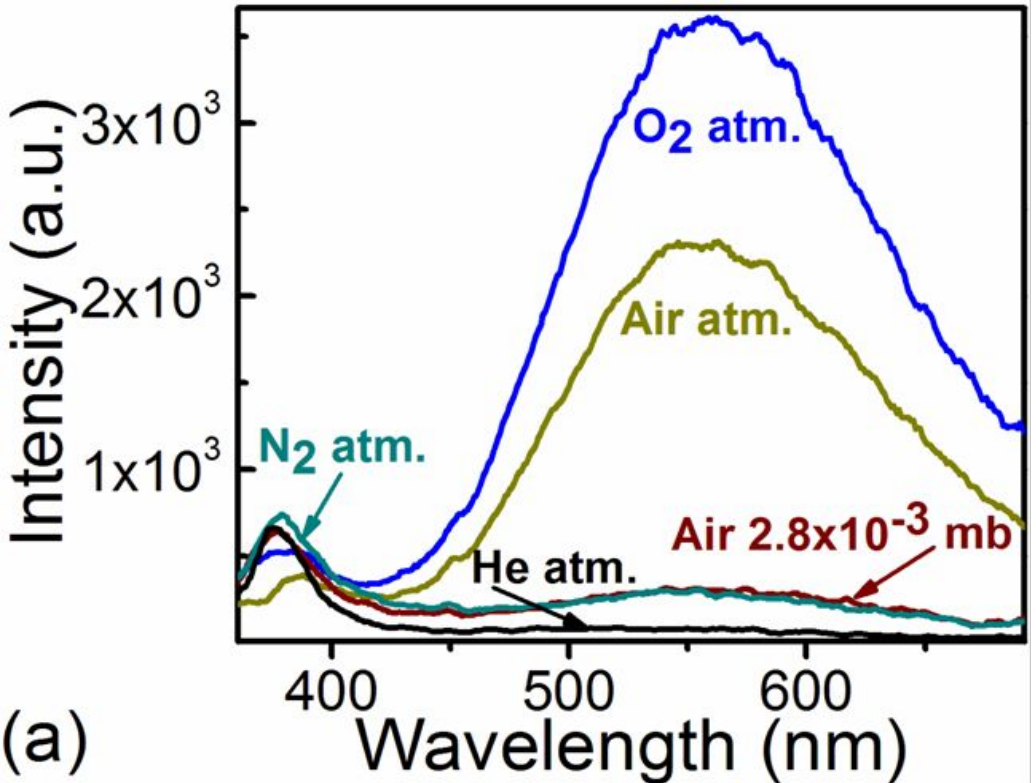

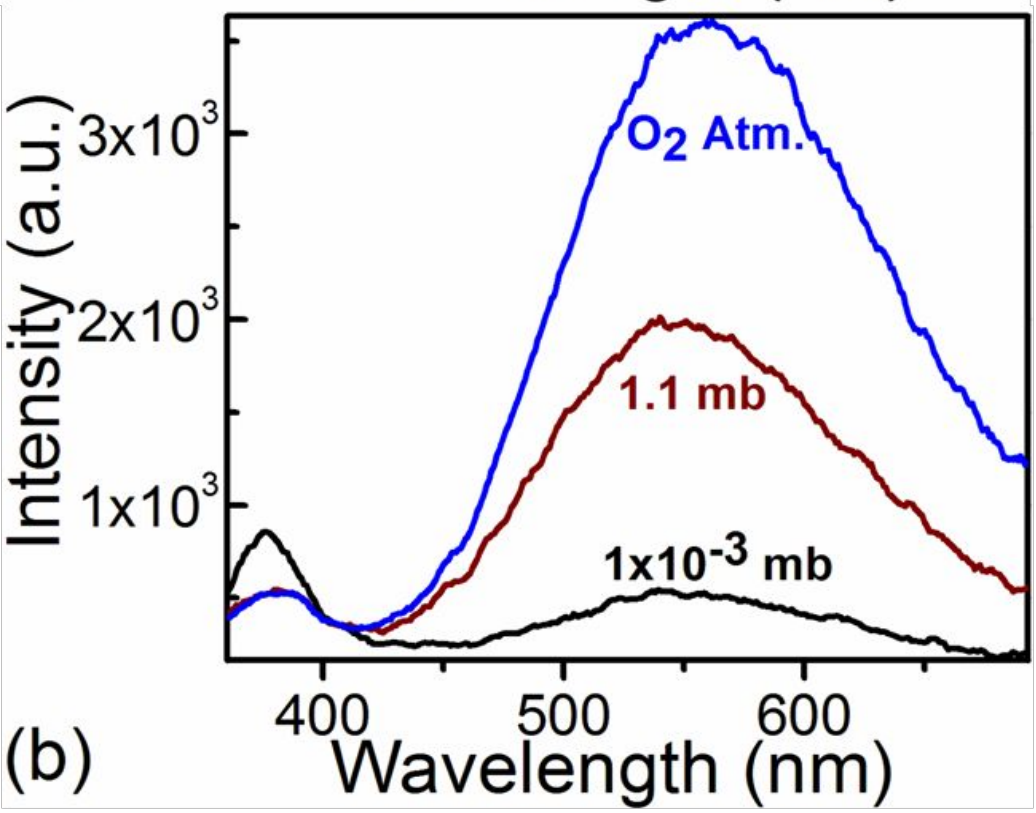

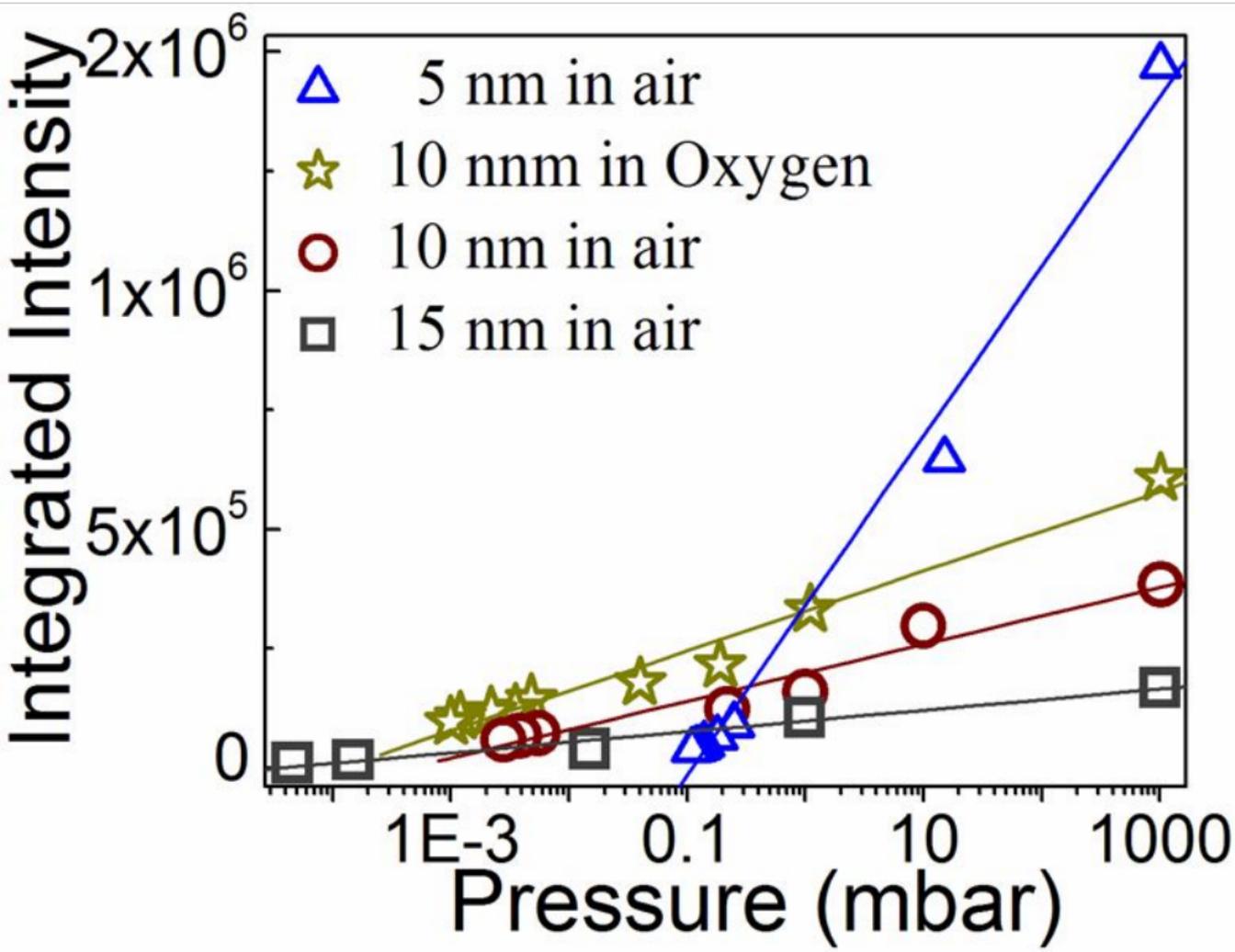

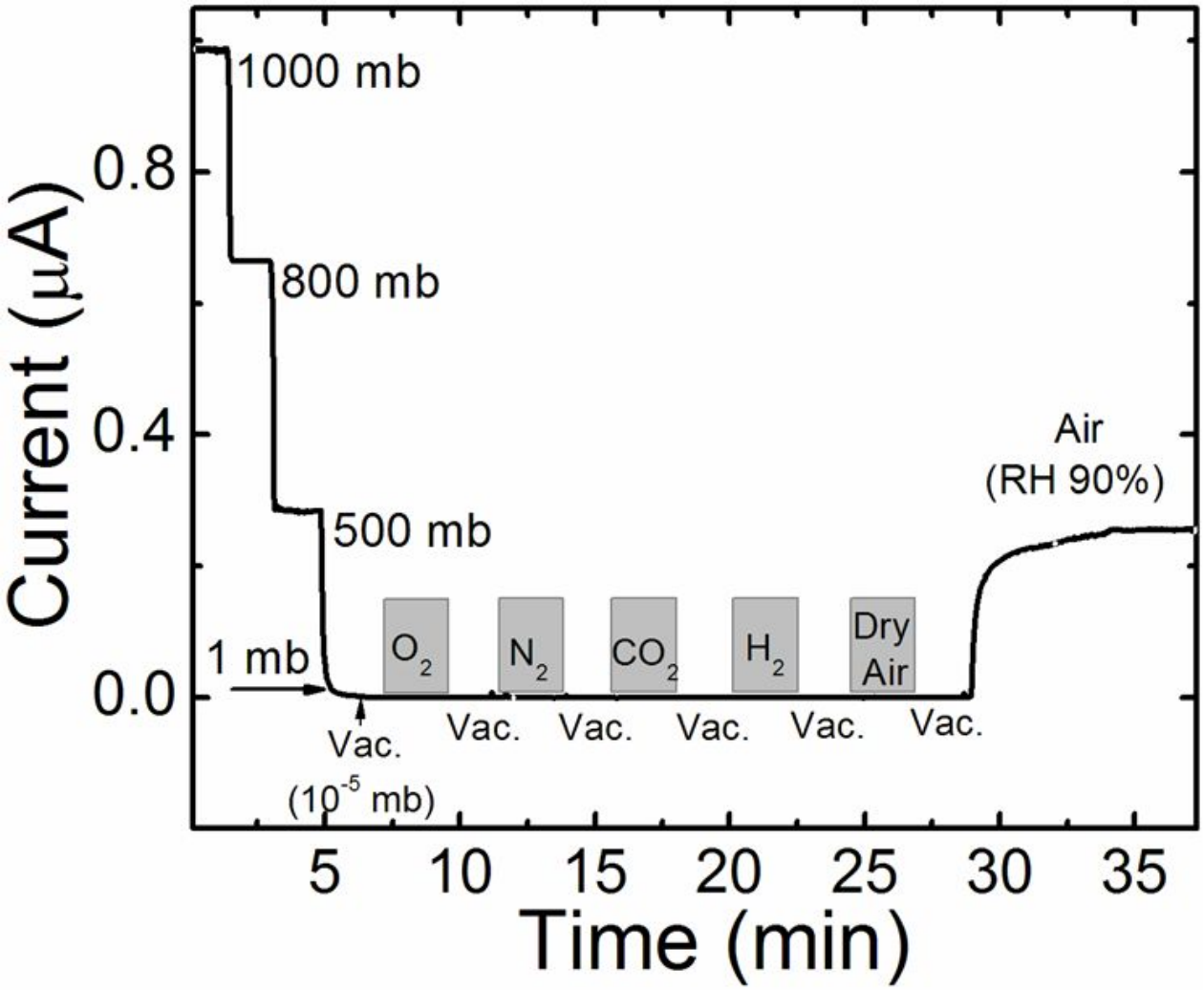

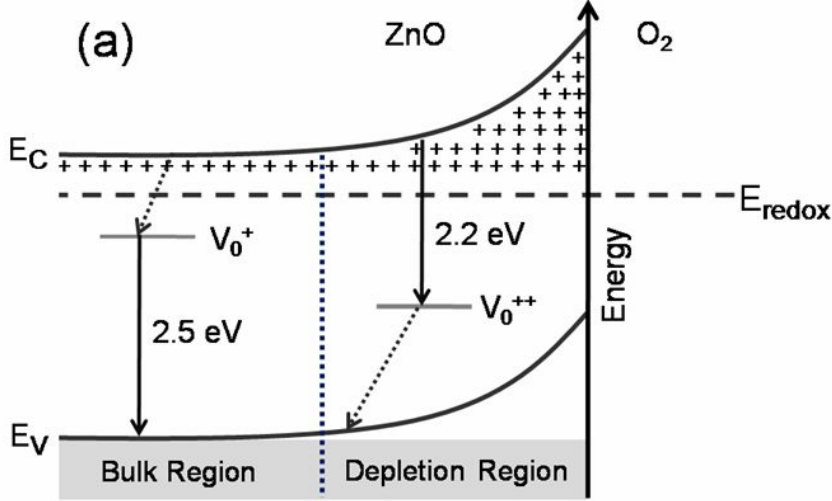

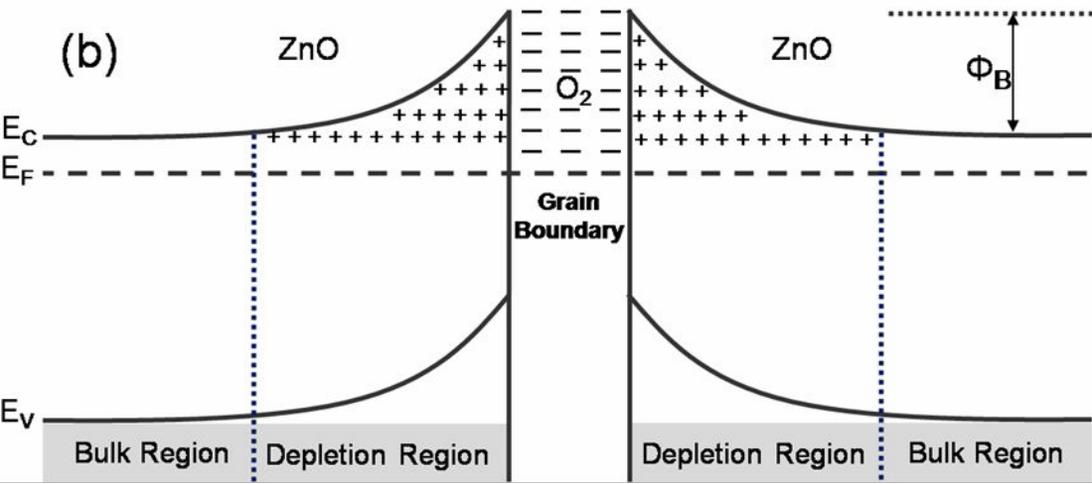